\documentclass[11pt,a4paper]{article}

\usepackage{bm}
\usepackage{amsmath}
\usepackage{algorithm}
\usepackage{algpseudocode}
\usepackage{import}
\usepackage{graphicx}
\usepackage{caption}
\usepackage{subcaption}
\usepackage{commath}
\usepackage{xcolor}
\usepackage{booktabs}
\usepackage[square,numbers]{natbib}
\usepackage{geometry}
\graphicspath{{./figures/}}

\begin{document}

\title{Fail\,-\,safe optimization of viscous dampers for seismic retrofitting}

\author{Nicol\`{o} Pollini\\nicolo@alumni.technion.ac.il}
\date{}



\maketitle

\begin{abstract}
\noindent This paper presents a new optimization approach for designing minimum-cost fail-safe distributions of fluid viscous dampers for seismic retrofitting. 
Failure is modeled as either complete damage of the dampers or partial degradation of the dampers' properties.
In general, this leads to optimization problems with large number of constraints. 
Thus, the use of a working-set optimization algorithm is proposed. 
The main idea is to solve a sequence of relaxed optimization sub-problems with a small sub-set of all constraints. 
The algorithm terminates once a solution of a sub-problem is found that satisfies all the constraints of the problem.
The retrofitting cost is minimized with constraints on the inter-story drifts at the peripheries of
frame structures.
The structures considered are subjected to a realistic ensemble of ground motions, and their response is evaluated with time-history analyses. 
The transient optimization problem is efficiently solved with a gradient-based sequential linear programming algorithm. The gradients of the response functions are calculated with a consistent adjoint sensitivity analysis procedure. Promising results attained for 3-D irregular frames are presented and discussed. The numerical results highlight the fact that the optimized layout and size of the dampers can change significantly even for moderate levels of damage.\\

\noindent \textit{keywords: fail-safe design, transient optimization, adjoint sensitivity analysis, viscous dampers, seismic retrofitting, passive control, earthquake engineering}
\end{abstract}


\maketitle


\section{Introduction}
\label{intro}
Fluid viscous dampers are a technology initially developed for military applications, but with the end of the Cold of War in 1990 their used was allowed also in civil engineering applications. 
They are in fact part of those technologies that were initially classified and their use restricted to the American military.
Because of their proven robustness and reliability during decades of Cold War applications, their use in commercial structures took place quickly \cite{taylor2019report}. 
In particular, the use of viscous dampers in earthquake engineering applications was first validated between the years 1990 and 1993, when it was also shown their benefit for wind and other types of transient excitation \cite{constantinou1992experimental,mcnamara2003fluid}.
Fluid viscous dampers are one of the passive energy dissipation devices available, and, broadly speaking, their purpose is to dissipate part of the input energy coming from an earthquake, thus reducing the deformation demand on the structure. As a consequence, if the dampers are properly sized and placed the structural damage can be significantly reduced. The use of passive energy dissipation devices has gained much attention in academia and practice, and the reader is referred to the comprehensive textbooks for more details \cite{soong1997passive,takewaki2011building,filiatrault2006principles}.

Two aspects strongly influence the structural performance of an added damping system made of fluid viscous dampers.
The first is the size of the dampers, which is typically expressed in terms of their damping coefficient. The second is the distribution of the dampers in the structure that needs to be retrofitted.
At the same time, these aspects affect not only the structural performance but also the associated retrofitting cost which can play a central role in promoting the use of fluid viscous dampers over other more traditional seismic retrofitting techniques. These aspects led to the development of several approaches for the sizing and placement of fluid viscous dampers assisted by optimization, as recently reviewed by De Domenico et al. \cite{de2019design}.
The available methodologies can be grouped based on the formulation used in the optimization problem.
A first group consists of those approaches that rely on continuous design variables (i.e. damping coefficients) \cite{gluck1996design,wu1997optimal,takewaki1997optimal,lavan2005optimal,lavan2006optimal,lavan2006boptimal,lavan2006optimalper,levy2006fully,bhaskararao2007optimum,silvestri2007physical,lavan2008noniterative,cimellaro2009integrated,almazan2009torsional,ribakov2011method,lavan2015methodology,altieri2018reliability}
In practice, the damping coefficients of the dampers are a continuous design variable that can be adapted to the needs of each specific project. 
This leads to an optimization problem computationally efficient and applicable also to large scale problems. 
However, it implies that the optimized design attained may consist of a wide variety of different damper sizes.
Each size-groups of dampers (i.e. dampers with the same mechanical properties) has specific costs associated to the production of that specific damper size and its prototype testing. 
Hence, the number of different size-groups of dampers is often limited to reduce some of the costs.
Thus, another group consists of methodologies that make use of discrete design variables to represent the damping coefficients and that lead to optimized designs of dampers characterized by a limited number of size-groups \cite{zhang1992seismic,agrawal1999optimal,dargush2005evolutionary,lavan2009multi,kanno2013damper}
These approaches rely on predefined parameters for the damping such as the available dampers’ sizes or the number of dampers, and this may have a considerable restraining effect on the optimized design solutions that can be attained. 
Moreover, in some of the cases referenced above the resulting optimization problems are relatively difficult to solve, compared to problems with continuous variables, due to the combinatorial nature of the resulting optimization problem.
Recently, an attempt was made to develop methodologies that combine the good aspects of the two set of approaches mentioned above: practical (i.e. near discrete) distributions of dampers with a reasonable computational cost \cite{lavan2014simultaneous,pollini2016towards,pollini2017minimum,pollini2018optimization}. 
This is achieved using a gradient-based continuous optimization approach for the placement and sizing of linear viscous dampers coupled with material interpolation techniques, typically applied in topology optimization\cite{bendsoe2013topology}. 
The dampers are selected from a limited number of available size-groups (whose properties are simultaneously optimized), and distributed in irregular 3-D frames by an optimization algorithm.
In this work, we rely on a continuous problem formulation as the focus of the work discussed herein is on the new fail-safe optimization approach for the design of viscous dampers. 

The optimization approaches discussed previously identify optimized dampers' distributions under a given set of conditions. 
These conditions are typically defined by the objective function to be minimized (e.g. the cost), the optimization constraints (e.g. the structural displacements or accelerations), and the type of modeling of the structural system considered.
Thus, for the given conditions considered the design solution attained is optimized and able to fulfill the performance objectives set by the designer.
Moreover, in all the optimization approaches listed above the obtained designs strongly rely on the fact that the dampers will perform as expected during an earthquake.
However, passive viscous damping systems may experience damage during their service lifetime or an earthquake, and therefore they may not perform as anticipated \cite{miyamoto2010limit}. 
This can potentially cause tragic human and economic losses.
To prevent an unexpected and undesired catastrophic failure of a structure retrofitted with viscous dampers, during the design phase fail-safe structural optimization methodologies should be applied.
The basic idea of the fail-safe structural design philosophy is that a structure should be designed to survive normal design loading conditions when damage occurs.
The resulting design is safe even if certain predefined types of damage conditions apply. 
The damage can be in the form of complete or partial failure of a structural member.
For the seismic retrofitting with viscous dampers, this can be translated into complete or partial failure of one or several dampers.

In the context of more traditional structural design applications, the fail-safe optimization for static loading has received an increasing attention by several researchers.
One of the first contributions is from 1976 by Sun et al. \cite{sun1976fail}, in the field of fail-safe truss structural optimization. The problems considered by Sun et al. minimize the structural weight with constraints on stresses, nodal displacements, and natural period of the structure. 
They define a priori the number and locations of damaged members.
In Achtziger and Bends{\o}e 1999 \cite{achtziger1999optimal} the structural topology optimization of degraded trusses is discussed.
The degradation is defined by a reduction of the modulus of elasticity in each structural member through a continuous variable.
The fail-safe structural topology optimization of continuum structures is first presented by Jansen et al. in 2014 \cite{jansen2014topology}, where the local failure of the structure is modeled by removing material in predefined portions of the 2-D design domain.
Zhou and Fleury \cite{zhou2016fail} in 2016 generalize the work of Jensen et al. \cite{jansen2014topology} to 3-D continuum structures by defining the position of the patches in a more refined way, relying in their formulation on spherical and cubic damage patches.
Kanno\cite{kanno2017redundancy} in 2017 proposes an approach for fail-safe structural optimization of trusses based on the worst-case scenario of structural degradation. The degradation is modeled by a predefined number of bars completely loosing their load carrying capacity.
L\"{u}deker and Kriegesmann \cite{ludeker2019fail} in 2019 discuss the fail-safe optimization of beam structures. The structural mass of the lattice structures considered is minimized with stress constraints. 
They define the failure scenarios as complete removal of one beam at a time. 
To reduce the computational cost during optimization, they propose a way of reducing the failure scenarios by combining the failures of certain beam elements into groups.
In recent work, Stolpe \cite{stolpe2019fail} discusses the fail-safe optimization of truss structures. 
Both partial and full failure of the truss elements is considered, and the optimization problem is formulated as a convex conic programming problem. 
Hence, the global optimal solution of the problem can be found. 
All the damage scenarios are considered, and to reduce the computational burden a working-set approach is used. 
It consists on cyclically solving sub-problems with expanding sub-sets of constraints until a solution of a sub-problem that satisfies all of the constrains is found.
It should be noted that the references listed above focus on the fail-safe structural optimization with static loading conditions.

Therefore, even though the attention towards fail-safe approaches for the structural optimization of static load bearing structures seems to be growing in recent years and to gain momentum, to the best of the author knowledge the fail-safe optimization of structural dynamics problems has not been discussed yet. 
This is also true in the context of seismic retrofitting with viscous dampers, where the use of a fail-safe approach would lead to optimized dampers' distributions with superior performance and levels of safety, something much needed in the earthquake engineering community. The purpose of this paper is to fulfill this need.

Thus, this paper presents a novel optimization approach for the minimum-cost fail-safe optimization of fluid viscous dampers for seismic retrofitting. 
The dampers' placement and size is simultaneously optimized. 
Failure is modeled as either complete or partial damage of the dampers. 
The damage is expressed though degradation of the mechanical properties of the dampers, i.e. their damping coefficient.
In general, this results in an optimization problem with a large number of constraints which requires a high computational effort during the optimization process.
The computational cost is reduced using a working-set strategy, similarly to Verbart and Stolpe 2018 \cite{verbart2018working}: a sequence of relaxed sub-problems each defined by an expanding sub-set of the constraints is solved cyclically. 
In every cycle new constraints that are active in correspondence of the current optimized solution are added, and the updated optimization sub-problem is solved again. 
The procedure stops once a solution that satisfies simultaneously all of the constraints of the original problem is found.
The dampers' retrofitting cost is minimized with constraints on the performance of the retrofitted structure. 
In this work the structural performance is measured in terms of inter-story drifts, or in other terms the relative displacements of the columns' ends at consecutive floors.
The response of the retrofitted structures are evaluated with time-history analyses considering an ensemble of realistic ground motions.
From a mathematical point of view, the problem at hand is nonlinear and nonconvex. 
It is formulated through continuous design variables (the damping coefficients of the dampers) and it is solved with a computationally efficient gradient-based algorithm, based on sequential linear programming. 
The gradients of the inter-story drift constraints are calculated consistently with an adjoint sensitivity analysis, based on a discretize-then-differentiate type of approach.

The reminder of the article is organized as follows: Section \ref{sec:dammod} presents the modeling of the damage scenarios, both for complete and partial damage of the dampers. Section \ref{sec:probform} describes the optimization problem formulation with details on the objective cost function, the governing equations for the structural dynamic equilibrium, and the response constraints. Section \ref{sec:workset} focuses on the proposed working-set strategy, used to reduce the computational cost during the fail-safe optimization process. The details of the adjoint sensitivity analysis and other computational considerations are given in Section \ref{sec:sensanal}. Numerical results are presented and discussed in Section \ref{sec:numex}, followed by concluding remarks in Section \ref{sec:final}.



\section{Failure scenarios}
\label{sec:dammod}
The problem considered herein consists in sizing and placing linear fluid viscous dampers in given frame structures subjected to realistic ground motions. The designs are identified by minimizing the dampers' cost while fulfilling expected performance criteria of the retrofitted structure. Moreover, the dampers' layouts are identified while also considering possible failure or damage scenarios thus obtaining safer optimized designs. 
Hence, in this section we start the discussion by presenting the scenarios considered for complete or partial damage of the viscous dampers. 

\subsection{Complete failure of dampers}
\label{subsec:compdamage}
We consider linear fluid viscous dampers. Hence, their force velocity behavior is formulated as follows:
\begin{equation}\label{eq:lindamp}
    f_{d}= c_{d} \Dot{d}
\end{equation}
where $\Dot{d}$ is the derivative in time of the relative displacement between the damper's ends; 
$c_{d}$ is the damping coefficient of the damper; 
and $f_{d}$ is the damper's output resisting force.
The design variables of the problem are the damping coefficients of the dampers.
If we imagine that we have the possibility of placing a maximum of $N_{d}$ dampers in a structure for retrofitting purposes, then $\mathcal{I}=\{1, \dots, N_{d} \}$ is the set of all the dampers' indices. 
Thus, the resulting damping matrix due to the added damping if calculated as follows:
\begin{equation}\label{eq:dampmat}
    \textbf{C}_{d} = \sum_{i \in \mathcal{I}} \textbf{T}^{T}_{i} c_{d,i} \textbf{T}_{i}
\end{equation}
where $c_{d,i}$ is the damping coefficient of the $i$-th damper; $\textbf{T}_{i}$ is a transformation matrix of the $i$-th damper from global coordinates to the local coordinates of the dampers (damper elongation); and $\textbf{C}_{d}$ is the added damping matrix.
We consider $n_{c}$ complete damage scenarios where in each case $m_{c}$ dampers are completely damaged. 
The set of completely damaged dampers' indices for the damage scenario $\alpha$ is $\mathcal{J}^{\alpha}\subseteq\mathcal{I}$.
The damaged added damping matrix is formulated as follows:
\begin{equation}\label{eq:dampmat2}
    \textbf{C}_{d}^{\alpha} = \sum_{i \in \mathcal{I} \setminus \mathcal{J}^{\alpha}} \textbf{T}^{T}_{i} c_{d,i} \textbf{T}_{i}
\end{equation}

\subsection{Partial failure of dampers}
\label{subsec:partdamage}
Similarly to the case of complete damage, we consider $n_{p}$ partial damage scenarios where in each case $m_{p}$ dampers are partially damaged. 
The set of partially damaged dampers' indices for the damage scenario $\alpha$ is $\mathcal{J}^{\alpha}\subseteq\mathcal{I}$.
The damaged added damping matrix is formulated as follows:
\begin{equation}\label{eq:dampmat3}
    \textbf{C}_{d}^{\alpha} = \sum_{i \in \mathcal{I} \setminus \mathcal{J}^{\alpha}} \textbf{T}^{T}_{i} c_{d,i} \textbf{T}_{i} + 
    \sum_{j \in \mathcal{J}^{\alpha}} \textbf{T}^{T}_{j} \nu_{j} c_{d,j} \textbf{T}_{j}
\end{equation}
where $\nu_{j}$ is a damage coefficient that reduces the damping capacity of the damper $j$, and that satisfies $0 \leq	\nu_{j} \leq1$ (e.g. $\nu_{j}=0.5\; \forall\; j$).


\section{Optimization problem formulation}
\label{sec:probform}
In this section we provide important details regarding the cost function minimized in the optimization analysis, the governing equations of motion, and the structural response constraints considered. We conclude this section presenting the final optimization problem formulation.

\subsection{Objective cost function and design variables}
\label{subsec:objvar}
The overall aim of this work is to propose a realistic and fail-safe optimization approach for minimizing the retrofitting cost associated to the seismic retrofitting with fluid viscous dampers.
In recent work, a realistic cost function for the retrofitting with viscous dampers has been proposed \cite{pollini2016towards}. The cost function includes the costs associated to the number of locations in the structure taken by dampers, the dampers' manufacturing cost, and the cost associated to the dampers' prototype testing.
In this work, the focus is on the novel fail-safe optimization approach for seismic retrofitting with viscous dampers proposed, and we will consider only the cost associated to the manufacturing of the dampers.
The cost of a single fluid viscous damper is a function of the peak force and stroke (maximum elongation) for which the damper is designed for.
The peak stroke is strongly correlated with the peak inter-story drift, which is constrained in our problem formulation. 
For this reason the damper stroke is not explicitly considered in the cost formulation here.
Assuming a dominant mode behavior, the velocity in the damper in location $i$ is proportional to $\omega_{1}d_{i}$, where $\omega_{1}$ is the dominant frequency and $d_{i}$ is the envelope peak drift at the location $i$. Experience shows that usually dampers are located where the drifts reach their allowable values, that are known values \cite{levy2006fully}. Thus the maximum velocities are known in advance and minimizing the damping coefficient is equivalent to minimizing the peak force. 
Based on these considerations, the retrofitting cost function minimized is:
\begin{equation}\label{eq:cost1}
J(\textbf{c}_{d}) = \sum_{i=1}^{N_{d}} c_{d,i}    
\end{equation}
where $\textbf{c}_{d}$ is a vector that collects the damping coefficients $c_{d,i}$ of the dampers.
The damping coefficient of each damper is formulated as follows:
\begin{equation}\label{eq:cdxi}
c_{d,i} = \bar{c}_{d} x_{i},\; \text{with:}\; 0\leq x_{i}\leq 1, \; i=1,\dots,N_{d}    
\end{equation}
In Eq. \eqref{eq:cdxi} $\bar{c}_{d}$ is the maximum allowable damping coefficient considered in the optimization analysis and it is defined a priori (e.g. $\bar{c}_{d}=150000 \frac{kNs}{m}$); and 
$x_{i}$ are the actual optimization design variables collected in the vector $\textbf{x}$.
Hence, the objective function can be normalized by the parameter $\bar{c}_{d}$ leading to:
\begin{equation}\label{eq:cost2}
J(\textbf{x}) = \sum_{i=1}^{N_{d}} x_{i}    
\end{equation}
Thus, the objective cost function $J(\textbf{x})$ of Eq. \eqref{eq:cost2} will be considered in the final optimization problem formulation.

\subsection{Equations of motion}
\label{subsec:eqmot}
In this work we consider linear structures equipped with linear fluid viscous dampers.
The seismic retrofitting of a structure is often performed with the goal of obtaining a linear behavior of the damped structure.
This can be achieved with the methodology proposed herein (as it is further discussed in Sec. \ref{subsec:respconstr}) by limiting the inter-story drifts to allowable limits that ensure a linear structural behavior, if feasible.
Moreover, we consider linear fluid viscous dampers because of their out-of-phase effect \cite{constantinou1992experimental}.
If needed the optimized distributions of linear dampers can be translated into equivalent nonlinear ones by equating the energy dissipated per cycle by linear and nonlinear dampers \cite{symans1998passive}.
Thus, the equations of motion for 3-D irregular structures considered herein are the following:
\begin{equation}\label{eq:eqmopt}
\begin{split}
& \textbf{M}\ddot{\textbf{u}}(t) + \left[\textbf{C}_{s}+\textbf{C}_{d}(\textbf{x})\right]\dot{\textbf{u}}(t) + \textbf{K}\textbf{u}(t)= - \textbf{M} \textbf{e}a_{g}(t)\\
& \textbf{u}(0) = \textbf{u}_{0}, \;\dot{\textbf{u}}(0) = \dot{\textbf{u}}_{0}
\end{split}
\end{equation}
where $\textbf{M}$, $\textbf{C}_{s}$, and $\textbf{K}$ are the mass, inherent damping, and stiffness matrices of the structure, respectively; 
$\textbf{C}_{d}$ is the added damping matrix that depends on the design variables $\textbf{x}$;
$\textbf{u}(t)$; $\dot{\textbf{u}}(t)$; and $\ddot{\textbf{u}}(t)$ are the displacements, velocities, and accelerations at time $t$ of the degrees of freedom relatively to the ground; 
$\textbf{e}$ is the influence vector and it represents the displacements of the masses resulting from static application of a unit ground displacement.
Basically, the vector $\textbf{e}$ assigns the acceleration to the degrees of freedom of the structure affected by the ground motion. 
$a_{g}(t)$ is the ground acceleration record as a function of time. 
In general, the local coordinates of the dampers are different from the global ones.
Thus, a transformation of coordinates is performed during the assembly of the added damping matrix $\textbf{C}_{d}$, as it is shown in Eq. \eqref{eq:dampmat}.

The equations of motion Eq. \eqref{eq:eqmopt} are discretized and solved in time using Newmark's time-stepping method \cite{chopra1995dynamics}. 
The particular approach adopted for evaluating the structural response directly affects the adjoint sensitivity analysis (that will be discussed in Sec. \ref{sec:sensanal}) where the gradients of the response constraints are calculated. 
Therefore for the sake of clarity, we provide the details of the time-stepping scheme adopted in this work.
The equations are discretized in time, and accelerations and velocities are expressed in terms of displacements:
\begin{equation}\label{eq:eqmopt2}
\begin{split}
&\textbf{M}\ddot{\textbf{u}}_{i+1} + \left[\textbf{C}_{s}+\textbf{C}_{d}(\textbf{x})\right]\dot{\textbf{u}}_{i+1} + \textbf{K}_{s}\textbf{u}_{i+1} = \textbf{P}_{i+1}\\
 \text{with: }&\textbf{P}_{i+1} = - \textbf{M} \textbf{e}a_{g,i+1}\\
 & \ddot{\textbf{u}}_{i+1} = \frac{1}{\beta \Delta t ^{2}} (\textbf{u}_{i+1}-\textbf{u}_{i})-\frac{1}{\beta \Delta t}\dot{\textbf{u}}_{i} - \left( \frac{1}{2 \beta}-1\right)\ddot{\textbf{u}}_{i}\\
 & \dot{\textbf{u}}_{i+1} = \frac{\gamma}{\beta \Delta t } (\textbf{u}_{i+1}-\textbf{u}_{i}) + \left( 1-\frac{\gamma}{\beta}\right)\dot{\textbf{u}}_{i} + \Delta t \left( 1-\frac{\gamma}{2 \beta}\right)\ddot{\textbf{u}}_{i}
\end{split}
\end{equation}
where $\beta=1/4$ and $\gamma=1/2$ are used in the average acceleration method, and $\beta=1/6$ and $\gamma=1/2$ are used in the linear acceleration method. Here we rely on the average acceleration method because it is stable for any choice of $\Delta t = t_{i+1}-t_{i}$.
It should be noted that in each time step $i$+1 Eq. \eqref{eq:eqmopt2} are solved for $\textbf{u}_{i+1}$, and then $\dot{\textbf{u}}_{i+1}$ and $\ddot{\textbf{u}}_{i+1}$ are calculated.

\subsection{Structural response constraints}
\label{subsec:respconstr}
The retrofitting designs are obtained by solving an optimization problem with nonlinear constraints imposed on the structural performance of the structure equipped with fluid viscous dampers.
In principle, there are several local responses of interest associated to the structural behavior and damage, such as inter-story drifts, total story accelerations, to name a few \cite{lavan2015optimal}.
Thus, in this work inter-story drifts are considered, which have been shown to be a good measure of both structural and nonstructural damage in many cases  \cite{charmpis2012optimized}.
In particular, the peak inter-story drifts normalized by a predefined allowed value are chosen as the local performance indices constrained in the optimization problem:
\begin{equation}\label{eq:constr1}
    d_{c,i} = \max_{t}(d_{i}(\textbf{x},t)/d_{allow}) \leq 1, \; \forall i=1,\ldots, N_{drifts}
\end{equation}
where $d_{i}(t)$ is the $i$-th inter-story drift constrained at time $t$; $d_{allow}$ is the maximum allowed value of inter-story drift (e.g. $3.5$ cm); 
$N_{drifts}$ is the number of inter-story drifts constrained.

The optimization approach adopted in this work relies on a gradient-based algorithm. Hence, all the objective and constrains functions involved in the optimization problem formulation have to be differentiable.
The constraint formulation of Eq. \eqref{eq:constr1} relies on the $\max$ function, which is not differentiable.
We use a p-norm function that approximates $\max_{t}(d_{i}(\textbf{x},t)/d_{allow})$ \cite{pollini2016towards} and that is differentiable:
\begin{equation}\label{eq:constr2}
    \Tilde{\textbf{d}}_{c}(\textbf{u},\textbf{x}) = \left( \frac{1}{t_{f}-t_{0}} \int _{t_{0}}^{t_{f}} \left( D^{-1}(\textbf{d}_{allow}) D(\textbf{H}\textbf{u}(\textbf{x},t))\right)^{p} dt \right)^{1/p} \cdot \textbf{1}  \leq \textbf{1}
\end{equation}
where $t_{0}$ and $t_{f}$ are the initial and final time of the time history analysis; $p$ is a large even number; $D()$ is an operator that transforms a vector into a diagonal matrix; 
$\textbf{H}$ is a matrix that transforms displacements $\textbf{u}$ into inter-story drifts $\textbf{d}$.
Additionally, in order to reduce the number of gradients of constraint functions from $N_{drifts}$ to one, we aggregate the constraints \eqref{eq:constr2} into a single constraint:
\begin{equation}\label{eq:constr3}
    g(\textbf{u},\textbf{x}) = \frac{\textbf{1}^{T} D(\Tilde{\textbf{d}}_{c}(\textbf{u},\textbf{x}))^{q+1} \textbf{1}}{\textbf{1}^{T}  D(\Tilde{\textbf{d}}_{c}(\textbf{u},\textbf{x}))^{q} \textbf{1}}-1 \leq 0
\end{equation}
Eq. \eqref{eq:constr3} is differentiable, and when $q$ is a large number Eq. \eqref{eq:constr3} approximates with more accuracy the maximum value of  $\Tilde{\textbf{d}}_{c}(\textbf{u},\textbf{x})$.

In this work we consider $n_{c}$ scenarios of total failure of $m_{c}$ dampers, and $n_{p}$ scenarios of partial failure of $m_{p}$ dampers. Additionally we consider also the case without any failure of the dampers.
Thus, in total for a given design layout of dampers we evaluate the structural response for $N_{FS}=1+n_{c}+n_{p}$ different scenarios. 
In Sec. \ref{sec:workset} we present a procedure used to reduce the number of failure scenarios actually considered during the optimization analysis to only a sub-set. 
However, in principle the number of responses constraints considered is $N_{FS}$:
\begin{equation}\label{eq:constr4}
    g_{\alpha}(\textbf{u},\textbf{x}) = \frac{\textbf{1}^{T} D(\Tilde{\textbf{d}}^{\alpha}_{c}(\textbf{u},\textbf{x}))^{q+1} \textbf{1}}{\textbf{1}^{T}  D(\Tilde{\textbf{d}}^{\alpha}_{c}(\textbf{u},\textbf{x}))^{q} \textbf{1}}-1 \leq 0, \; \text{for } \alpha=1,\dots,N_{FS}
\end{equation}
Eq. \eqref{eq:constr4} represents a list of $N_{FS}$ nonlinear constraints considered in the optimization analysis. It should be noted that in principle $N_{FS}$ can be a large number, resulting in lot of response and sensitivity analyses that need to be performed in every optimization iteration.

\subsection{Final optimization problem}
\label{subsec:optprob1}
We present now the final optimization problem formulation for the fail-safe design of fluid viscous dampers for seismic retrofitting.
The optimization problem is stated as follows:
\begin{equation}\label{eq:optprob}\tag{$\mathcal{P}_{FS}$}
\begin{split}
 \underset{\textbf{x} \in {\rm I\!R}^{N_{d}}}{\text{minimize:}} & \quad J(\textbf{u},\textbf{x}) \\
 \text{subject to: } & \quad g_{\alpha}(\textbf{u},\textbf{x})  \leq 0, \; \forall \; \alpha \in \mathcal{S}_{FS}\\
& \quad 0\leq x_{i}\leq 1,\; \text{for } i=1,\dots, N_{d}\\
\text{with: } &\quad \textbf{M}\ddot{\textbf{u}}(t) + \left[\textbf{C}_{s}+\textbf{C}_{d}(\textbf{x})\right]\dot{\textbf{u}}(t) + \textbf{K}\textbf{u}(t)= - \textbf{M} \textbf{e}a_{g}(t)\quad \forall a_{g}\in\mathcal{E}\\
&\quad \textbf{u}(0) = \textbf{u}_{0}, \;\dot{\textbf{u}}(0) = \dot{\textbf{u}}_{0}
\end{split}
\end{equation}
where $\mathcal{E}$ is the ensemble of ground motions considered. $\mathcal{S}_{FS}$ is the set of all indices that identify the failure scenarios considered, and its cardinality is $|\mathcal{S}_{FS}|=N_{FS}$.
The optimization problem \ref{eq:optprob} is solved with a sequential linear programming approach. 
With this approach, in every optimization iteration the problem is linearized and solved locally. 
Hence, the gradients of the objective and constraints functions need to be calculated. 
The objective function $J$ is formulated explicitly in terms of the design variables of the problem. Thus, its gradient can be calculated directly. The gradients of the constraint functions need to be calculated with a dedicated adjoint sensitivity analysis instead. More details about the sensitivity analyses performed in this work are given in Sec. \ref{sec:sensanal}, but from a computational point of view each sensitivity analysis weights as much as a linear response time-history analysis. In principle, having $N_{FS}$ fail-save scenarios means that in each optimization iteration $N_{FS}$ adjoint sensitivity analyses need to be performed in order to calculated the gradients of the $ g_{\alpha}(\textbf{u},\textbf{x})$ constraint functions, for $\alpha=1,\dots,N_{FS}$. Thus, in order to reduce the number of sensitivity analyses performed in each optimization iteration, and hence the overall computational cost, in Sec. \ref{sec:workset} we propose a working set- strategy. The goal is to consider a subset of constraints during the optimization, and to reduce in this way the required computational cost and time.


\section{Working-set strategy}
\label{sec:workset}
In the fail-safe optimization approach discussed herein, we consider $N_{FS}$ failure scenarios.
In every optimization iteration, the structural response should be evaluated for each of these scenarios.
Moreover, the gradient of the aggregated drift constraint \eqref{eq:constr4} is calculated for each failure scenario in each optimization iteration as well, at the cost of an additional time-history analysis per constraint (more details are given in Sec. \ref{sec:sensanal}).
Thus, the solution of \eqref{eq:optprob} requires in  every iteration the evaluation of $2 \times N_{FS}$ time-history analyses, and this may result in a very high computational cost.
Thus, in this section we present the working-set strategy adopted in order to reduce the number of failure scenarios (hence constraints) actually considered in the optimization analysis, and as a consequence the overall computational cost. This strategy is inspired by the one presented by Verbart and Stolpe  \cite{verbart2018working}.
We present the relaxed formulation of the optimization sub-problems considered, the strategy for updating the set of constraints considered between consecutive sub-problems $k$ and $k+1$, and the stopping criterion for the optimization process.

In particular, instead of solving the full optimization problem \ref{eq:optprob}, we solve a sequence of relaxed sub-problems that consider a sub-set of failure scenarios:
\begin{equation}\label{eq:optprob2}\tag{$\mathcal{P}_{WS}^{k}$}
\begin{split}
 \underset{\textbf{x} \in {\rm I\!R}^{N_{d}}}{\text{minimize:}} & \quad J(\textbf{u},\textbf{x}) \\
 \text{subject to: } & \quad g_{\alpha}(\textbf{u},\textbf{x})  \leq 0, \; \forall \; \alpha \in \mathcal{S}_{WS}^{k}\\
& \quad 0\leq x_{i}\leq 1,\; \text{for } i=1,\dots, N_{d}\\
\text{with: } &\quad \textbf{M}\ddot{\textbf{u}}(t) + \left[\textbf{C}_{s}+\textbf{C}_{d}(\textbf{x})\right]\dot{\textbf{u}}(t) + \textbf{K}\textbf{u}(t)= - \textbf{M} \textbf{e}a_{g}(t)\quad \forall a_{g}\in\mathcal{E}\\
&\quad \textbf{u}(0) = \textbf{u}_{0}, \;\dot{\textbf{u}}(0) = \dot{\textbf{u}}_{0}
\end{split}
\end{equation}
where $\mathcal{S}^{k}_{WS}\subset\mathcal{S}_{FS}$ is the sub-set of indices that identify the failure scenarios actually considered in the $k$-th sub-problem. Its cardinality is $|\mathcal{S}_{WS}^{k}|=N_{WS}^{k}$, with  $N_{WS}^{k}<<N_{FS}$. Once the sub-problem $\mathcal{P}^{k}_{WS}$ has been solved, new active constraints are added to the following sub-problem $\mathcal{P}^{k+1}_{WS}$.

A working set $\mathcal{S}^{k}_{WS}$ is a set of constraints indices considered while solving the optimization sub-problem $\mathcal{P}^{k}_{WS}$. 
The first working-set considered, $\mathcal{S}^{k}_{WS}$ with $k=0$, does not include any failure scenarios. 
It contains only one aggregated drift constraint, as in Eq. \eqref{eq:constr3}, associated to the case without any failure.
This leads to a dampers' design that does not account for any failure scenarios, similarly to the approach of Lavan and Levy \cite{lavan2006optimalper}. We denote this solution as $\textbf{x}_{0}$. 
Then, all failure constraints are checked against this solution, and the most critical ones are included in the first working-set $\mathcal{S}^{k+1}_{WS}$: 
\begin{equation}\label{eq:firstws}
\mathcal{T}^{k}=\{ i\; |\; (g^{k}_{max}-g^{k}_{i})/ g^{k}_{max} \leq \epsilon\}
\end{equation}
where $\mathcal{T}^{k}$ is a temporary set that contains the indices of failure scenarios not included in $\mathcal{S}^{k}_{WS}$, and $g^{k}_{max}$ is the maximum constraint value:
\begin{equation}
g^{k}_{max}=\underset{i}{\max}\left(g^{k}_{i}\right) \text{ for } i \in \mathcal{S}_{FS}
\end{equation}
The parameter $\epsilon$ determines the number of constraints considered as critical and hence included in the working-set $\mathcal{T}^{k}_{WS}$ (e.g. $\epsilon=5\%$). The new working-set $\mathcal{S}^{k+1}_{WS}$ is then defined as:
\begin{equation}\label{eq:updatews}
\mathcal{S}^{k+1}_{WS} = \mathcal{T}^{k+1} \cup \mathcal{S}^{k}_{WS}
\end{equation}
and it is updated after solving the optimization sub-problem $\mathcal{P}_{WS}^{k+1}$ based on the solution $\textbf{x}_{k+1}$.
The proposed strategy is built in a way that ensures that critical failure scenarios can only be added, and hence: $\mathcal{S}^{k}_{WS} \subset \mathcal{S}^{k+1}_{WS}$
As a result, it allows for significant computational savings both in terms of response analyses, and adjoint sensitivity analyses. 

The sequential solution of optimization sub-problems $\left\{ \mathcal{P}_{WS}^{k};\mathcal{P}_{WS}^{k+1};\dots \right\}$ is terminated once an optimized solution is found such that in correspondence of this solution the set of constraints in $\mathcal{S}_{FS}$ that are violated, namely $\mathcal{V}^{k}$, is empty:
\begin{equation}\label{eq:violate}
\mathcal{V}^{k}=\left\{ i \in \mathcal{S}_{FS} \; | \; g_{i} > 0 \right\}
\end{equation}
The pseudo code of the working-set strategy adopted herein is provided in Algorithm \ref{alg:algo1}.
\begin{algorithm}\caption{Working-set strategy as a sequence of optimization sub-problems} \label{alg:algo1}
\begin{algorithmic}[1]
\State Set: $k=0$; $|\mathcal{S}^{0}_{WS}|=1$ (non-failure scenario only); $flag=1$
\While{$flag=1$} 
\State{Solve $\mathcal{P}_{WS}^{k}$}
\State{Evaluate $\mathcal{V}^{k}$ as in Eq. \eqref{eq:violate}}
\If{$\mathcal{V}^{k}=\emptyset$} 
\State{$flag=0$ (i.e. stop)}
\EndIf
\State{Define $\mathcal{S}^{k+1}_{WS}$ based on Eq. \eqref{eq:updatews}}
\State{$k=k+1$}
\EndWhile
\end{algorithmic}
\end{algorithm}



\section{Sensitivity analysis and computational considerations}
\label{sec:sensanal}
As it has been already anticipated in Sec. \ref{subsec:optprob1}, the optimization sub-problems $\mathcal{P}_{WS}^{k}$ are solved with a modified sequential linear programming (SLP) approach inspired by the cutting planes method \cite{cheney1959newton,kelley1960cutting}.
This is an iterative approach, where in every optimization iteration the problem is linearized and solved locally.
Hence, the gradients of the objective and constraints functions need to be calculated. 
The objective function $J$ is formulated explicitly in terms of the design variables of the problem. Thus, its gradient $\bm{\nabla} J$ can be calculated directly: $\bm{\nabla} J=\textbf{1}$, where $\textbf{1}$ is a vector with all entries equal to one and dimensions $[N_{d} \times 1]$.
The gradient of each aggregated constraint (i.e. $\bm{\nabla} g_{\alpha}$), on the other hand, requires a sensitivity analysis.
Since we assume that the number of design variables is larger than the number of constraints considered in each sub-problem, we rely on an adjoint sensitivity analysis \cite{michaleris1994tangent}. 
This is ensured by the working-set strategy adopted, which has been described in Sec. \ref{sec:workset}.
In the case of a number of design variables smaller than the number of constraints, it would have been recommended to adopt the direct differentiation method for calculating the constraints' gradients \cite{verbart2018working}.
Moreover, to ensure the consistency of the sensitivity calculated we rely on the so called discretize-then-differentiate adjoint variable method \cite{le2012material,jensen2014consistency,pollini2018adjoint,lavan2019adjoint}. According to this method, the discrete version of the governing equilibrium equations \eqref{eq:eqmopt2} is considered in the gradient calculation.

\subsection{Adjoint sensitivity analysis}
\label{subsec:adjsensanal}

The goal is to calculate the gradient of the constraint function defined in Eq. \eqref{eq:constr4}:
\begin{equation}
\bm{\nabla} g = \frac{d\textbf{u}}{d\textbf{x}}\frac{dg}{d\textbf{u}}
\end{equation}
where for simplicity we have dropped the subscript $\alpha$.
First we define an augmented function $\hat{g}$, which is obtained by adding zero terms to the definition of $g$.
These terms are the residuals of the discrete dynamic equilibrium equations defined in Eq. \eqref{eq:eqmopt2}:
\begin{equation}
\hat{g}(\textbf{u},\textbf{x}) = g(\textbf{u},\textbf{x}) 
+ \sum_{i=1}^{N} \bm{\lambda}^{T}_{u,i} \textbf{R}_{u,i}
+ \sum_{i=1}^{N} \bm{\lambda}^{T}_{v,i} \textbf{R}_{v,i}
+ \sum_{i=1}^{N} \bm{\lambda}^{T}_{a,i} \textbf{R}_{a,i}
\end{equation}
where $N$ is the number of time steps; $\bm{\lambda}^{T}_{u,i}$, $\bm{\lambda}^{T}_{v,i}$, and $\bm{\lambda}^{T}_{a,i}$ are vectors that collect the adjoint variables; and:
\begin{equation}
\begin{split}
&\textbf{R}_{u,i} = \textbf{M}\ddot{\textbf{u}}_{i} + \left[\textbf{C}_{s}+\textbf{C}_{d}(\textbf{x})\right]\dot{\textbf{u}}_{i} + \textbf{K}_{s}\textbf{u}_{i} + \textbf{M}\textbf{e}a_{g,i} \\
 & \textbf{R}_{v,i}= -\dot{\textbf{u}}_{i} + \frac{\gamma}{\beta \Delta t } (\textbf{u}_{i}-\textbf{u}_{i-1}) + \left( 1-\frac{\gamma}{\beta}\right)\dot{\textbf{u}}_{i-1} + \Delta t \left( 1-\frac{\gamma}{2 \beta}\right)\ddot{\textbf{u}}_{i-1}\\
 & \textbf{R}_{a,i}= -\ddot{\textbf{u}}_{i} + \frac{1}{\beta \Delta t ^{2}} (\textbf{u}_{i}-\textbf{u}_{i-1})-\frac{1}{\beta \Delta t}\dot{\textbf{u}}_{i-1} - \left( \frac{1}{2 \beta}-1\right)\ddot{\textbf{u}}_{i-1}
\end{split}
\end{equation}
When the equilibrium is satisfied in every time-step we have that $\hat{g}(\textbf{u},\textbf{x}) = g(\textbf{u},\textbf{x})$  and hence $\bm{\nabla}\hat{g}=\bm{\nabla} g$.
The gradient of the augmented function is then calculated as follows:
\begin{equation}\label{eq:discrdiff1}
\begin{split}
\bm{\nabla}\hat{g} = 
&\sum_{i=1}^{N} \left(\frac{d\textbf{u}_{i}}{d\textbf{x}}\frac{dg}{d\textbf{u}_{i}} + \frac{d\dot{\textbf{u}}_{i}}{d\textbf{x}}\frac{dg}{d\dot{\textbf{u}}_{i}} +\frac{d\ddot{\textbf{u}}_{i}}{d\textbf{x}}\frac{dg}{d\ddot{\textbf{u}}_{i}} + \frac{dg}{d\textbf{x}} \right)\\
+ &\sum_{i=1}^{N}  \left( \frac{d\textbf{u}_{i}}{d\textbf{x}} \frac{d\textbf{R}_{u,i}}{d\textbf{u}_{i}} 
+  \frac{d\dot{\textbf{u}}_{i}}{d\textbf{x}} \frac{d\textbf{R}_{u,i}}{d\dot{\textbf{u}}_{i}} 
+   \frac{d\ddot{\textbf{u}}_{i}}{d\textbf{x}} \frac{d\textbf{R}_{u,i}}{d\ddot{\textbf{u}}_{i}} + \frac{d\textbf{R}_{u,i}}{d\textbf{x}} \right) \bm{\lambda}_{u,i}\\
+ &\sum_{i=1}^{N}  \left( \frac{d\textbf{u}_{i}}{d\textbf{x}} \frac{d\textbf{R}_{v,i}}{d\textbf{u}_{i}} 
+  \frac{d\dot{\textbf{u}}_{i}}{d\textbf{x}} \frac{d\textbf{R}_{v,i}}{d\dot{\textbf{u}}_{i}} 
+   \frac{d\ddot{\textbf{u}}_{i}}{d\textbf{x}} \frac{d\textbf{R}_{v,i}}{d\ddot{\textbf{u}}_{i}} + \frac{d\textbf{R}_{v,i}}{d\textbf{x}}\right) \bm{\lambda}_{v,i}\\
+ &\sum_{i=1}^{N}  \left( \frac{d\textbf{u}_{i}}{d\textbf{x}} \frac{d\textbf{R}_{a,i}}{d\textbf{u}_{i}} 
+  \frac{d\dot{\textbf{u}}_{i}}{d\textbf{x}} \frac{d\textbf{R}_{a,i}}{d\dot{\textbf{u}}_{i}} 
+   \frac{d\ddot{\textbf{u}}_{i}}{d\textbf{x}} \frac{d\textbf{R}_{a,i}}{d\ddot{\textbf{u}}_{i}} + \frac{d\textbf{R}_{a,i}}{d\textbf{x}} \right) \bm{\lambda}_{a,i}
\end{split}
\end{equation}

To avoid the calculation of the implicit derivatives of the state variables with respect to the design variables (i.e. $\frac{d\textbf{u}_{i}}{d\textbf{x}}$, $\frac{d\dot{\textbf{u}}_{i}}{d\textbf{x}}$, and  $\frac{d\ddot{\textbf{u}}_{i}}{d\textbf{x}}$), once \eqref{eq:discrdiff1} is differentiated we collect all the terms multiplying these derivatives and we equate them to zero. 
Therefore, in each time-step $i=1,\dots, N-1$ we have that:
\begin{equation}\label{eq:discrdiff2}
\begin{split}
& \textbf{M}^{T} \bm{\lambda}_{u,i} + \bm{\lambda}_{a,i} - \Delta t \left( 1-\frac{\gamma}{2 \beta} \right) \bm{\lambda}_{v,i+1}  + \left( \frac{1}{2 \beta} -1\right)\bm{\lambda}_{a,i+1}=0\\
& \textbf{C}^{T} \bm{\lambda}_{u,i} +\bm{\lambda}_{v,i} -  \left( 1-\frac{\gamma}{\beta} \right) \bm{\lambda}_{v,i+1} +  \frac{1}{\beta\Delta t} \bm{\lambda}_{a,i+1}=0\\
& \textbf{K}^{T} \bm{\lambda}_{u,i}  - \frac{\gamma}{\beta \Delta t}  \left( \bm{\lambda}_{v,i}-\bm{\lambda}_{v,i+1} \right) - \frac{\gamma}{\beta \Delta t^{2} } \left( \bm{\lambda}_{a,i}-\bm{\lambda}_{a,i+1} \right) + \frac{dg}{d\textbf{u}_{i}}=0\\
\end{split}
\end{equation}
where $C=C_{s}+C_{d}(\textbf{x})$, and we also included the fact that $\frac{dg}{d\dot{\textbf{u}}_{i}}=0$, and $\frac{dg}{d\ddot{\textbf{u}}_{i}}=0$. In matrix form, the system of equations \eqref{eq:discrdiff2} can be written as $\textbf{A}\,\bm{\xi}_{i}=\textbf{b}_{i}$, where:
\begin{equation}\label{eq:discrdiff3}
\underbrace{\begin{bmatrix}
\textbf{M}^{T} & \textbf{0} & \textbf{I}\\
\textbf{C}^{T} & \textbf{I} & \textbf{0}\\
\textbf{K}^{T} & -\frac{\gamma}{\beta \Delta t}\textbf{I} & -\frac{\gamma}{\beta \Delta t^{2} }\textbf{I}\\
\end{bmatrix}}_{\textbf{A}}
\underbrace{\begin{bmatrix}
\bm{\lambda}_{u,i}\\
\bm{\lambda}_{v,i}\\
\bm{\lambda}_{a,i}
\end{bmatrix}}_{\bm{\xi}_{i}}
=
\underbrace{\begin{bmatrix}
\Delta t \left( 1-\frac{\gamma}{2 \beta} \right)\bm{\lambda}_{v,i+1}-\left( \frac{1}{2 \beta} -1\right)\bm{\lambda}_{a,i+1}\\
\left( 1-\frac{\gamma}{\beta} \right)\bm{\lambda}_{v,i+1}-\frac{1}{\beta\Delta t}\bm{\lambda}_{a,i+1}\\
-\frac{\gamma}{\beta \Delta t}\bm{\lambda}_{v,i+1} - \frac{\gamma}{\beta \Delta t^{2} }\bm{\lambda}_{a,i+1}-\frac{dg}{d\textbf{u}_{i}}
\end{bmatrix}}_{\textbf{b}_{i}}
\end{equation}
where $\textbf{I}$ is the identity matrix with dimensions $[N_{dof}\times N_{dof}]$; $N_{dof}$ is the number of structural degrees of freedom; $\textbf{A}$ has dimensions $[3N_{dof}\times 3N_{dof}]$, $\bm{\xi}_{i}$ and $\textbf{b}_{i}$ $[3N_{dof}\times 1]$.
Numerically, Eq. \eqref{eq:constr2} in calculated in discrete form as follows:
\begin{equation}\label{eq:constr5}
    \Tilde{\textbf{d}}_{c}(\textbf{u},\textbf{x}) = \left( \frac{1}{t_{f}-t_{0}} \sum _{i=1}^{N} w_{i}\left(  D^{-1}(\textbf{d}_{allow}) D(\textbf{H}\textbf{u}_{i}(\textbf{x}))\right)^{p} \right)^{1/p} \cdot \textbf{1}  
\end{equation}
where $w_{i}$ is a weight used for numerical integration (e.g. $w_{i}=\Delta t$ for $i\neq \{0,N\}$, and  $w_{i}=\Delta t/2$ for $i= \{0,N\}$). 
Thus, in Eq. \eqref{eq:discrdiff3} $\frac{dg}{d\textbf{u}_{i}}$ is explicitly calculated as follows:
\begin{equation}\label{eq:dgdu}
\begin{split}
\frac{dg}{d\textbf{u}_{i}} = &- \textbf{H}^{T} D^{-1}(\textbf{d}_{allow}) \left( \frac{1}{t_{f}-t_{0}} \sum _{i=1}^{N} \left( w_{i} D^{-1}(\textbf{d}_{allow}) D(\textbf{H}\textbf{u}_{i}(\textbf{x}))\right)^{p} \right)^{\frac{1-p}{p}}\\
& \frac{1}{t_{f}-t_{0}} w_{i} \left(D^{-1}(\textbf{d}_{allow}) D(\textbf{H}\textbf{u}_{i}(\textbf{x}))\right)^{p-1}\\
& \frac{1}{\left( den\right)^{2}}\left( den\, (q+1)D\left(\Tilde{\textbf{d}}_{c}\right)^{q}\textbf{1}-num\,q\,D\left(\Tilde{\textbf{d}}_{c}\right)^{q-1} \textbf{1}\right)
\end{split}
\end{equation}
where
\begin{equation}
num= \textbf{1}^{T} D\left(\Tilde{\textbf{d}}_{c}\right)^{q+1} \textbf{1}; \; den =\textbf{1}^{T} D\left(\Tilde{\textbf{d}}_{c}\right)^{q} \textbf{1}
\end{equation}
Essentially, the adjoint sensitivity analysis consists in solving for each time-step $i\rightarrow i-1$ the linear system of equations Eq. \eqref{eq:discrdiff3}. 
It results in a linear time-history analysis solved backwards in time, and with known final conditions:
\begin{equation}
\textbf{A}\,\bm{\xi}_{N}=\textbf{b}_{N} \text{ with } 
\textbf{b}_{N}=\begin{bmatrix}
\textbf{0}&
\textbf{0}&
-\frac{dg}{d\textbf{u}_{N}}
\end{bmatrix}^{T}
\end{equation}
It should be noted that the system of equations of the adjoint sensitivity analysis \eqref{eq:discrdiff3} has dimensions $[3N_{dof}\times 3N_{dof}]$. 
The dimensions of the system of equations for the evaluation of the structural response \eqref{eq:eqmopt2} are  $[N_{dof}\times N_{dof}]$, once the Newmark's method is implemented.
However, also the dimensions of the system of equations \eqref{eq:discrdiff3} can be reduced to $[N_{dof}\times N_{dof}]$ if, for example, $\bm{\lambda}_{v,i}$ and $\bm{\lambda}_{a,i}$ are expressed in terms of $\bm{\lambda}_{u,i}$.

Once the adjoint variables $\bm{\xi}_{i}$ are known in each time-step $i$, the constraint gradient is calculated as follows:
\begin{equation}\label{eq:finalgrad}
\bm{\nabla} g = \sum _{i=1}^{N} \dot{\textbf{u}}^{T}_{i} \, \frac{d C_{d}}{d \textbf{x}} \, \bm{\lambda}_{u,i}
\end{equation}
The pseudo code for the adjoint sensitivity analysis is provided in Algorithm \ref{alg:algo2}.
\begin{algorithm}\caption{Adjoint sensitivity analysis} \label{alg:algo2}
\begin{algorithmic}[1]
\State{Assemble the matrix $\textbf{A}$}
\State{Calculate the matrix $\textbf{A}^{-1}$}
\State{Calculate the final conditions $\textbf{b}_{N}$}
\State{Calculate $\bm{\xi}_{N}=\textbf{A}^{-1}\textbf{b}_{N}$}
\For{i=N-1,\dots,1}
\State{Calculate $\textbf{b}_{i}(\bm{\xi}_{i+1})$}
\State{Calculate $\bm{\xi}_{i}=\textbf{A}^{-1}\textbf{b}_{i}$}
\EndFor
\State{Initialize $\bm{\nabla} g = \textbf{0}$}
\For{i=1, \dots,N}
\State{$\bm{\nabla} g = \bm{\nabla} g + \dot{\textbf{u}}^{T}_{i} \, \frac{d C_{d}}{d \textbf{x}} \, \bm{\lambda}_{u,i}$}
\EndFor
\end{algorithmic}
\end{algorithm}


\subsection{Computational considerations}
\label{subsec:compcons}
In order to successfully adopt existing algorithms for the nonlinear and nonconvex optimization problems at hand, i.e. \eqref{eq:optprob2}, some practical and conservative measures need to be taken in the optimization algorithm implementation. These include: the selection of a dominant ground motion from the ensemble of records considered; the management of the linearized drift constraints; a continuation scheme for the control of certain parameters; and convergence criteria. \\

\noindent \textbf{Ground motion selection}.
In general, the full set of ground motions in the ensemble should be considered when solving \eqref{eq:optprob2}. However, this may further increase the computational cost. Thus, in this work we follow the procedure suggested by Lavan and Levy \cite{lavan2005optimal} where the dominant ``active'' ground motion is selected (i.e. the one with largest spectral displacements in correspondence to the structure’s natural period). Once the optimization with the single ground motion is terminated, the optimized design is tested with all the acceleration records from the ensemble and the ground motions for which the drift constraint is violated are added. The process terminates when the drift constraint is not violated with all the ground motions.\\

\noindent \textbf{Managing the linearized constraints}.
As is has been already mentioned, we apply a modified SLP approach inspired by the cutting planes method to solve \eqref{eq:optprob2}. In every iteration of standard SLP, a linear subproblem is solved. In the algorithm used herein, the sub-problems grow in dimension, because in each iteration a new linearized approximation of the aggregated constraint (one for each $\alpha \in \mathcal{S}_{WS}^{k}$) is added to the set of constraints considered . Because the problem at hand is non-convex, it may happen that a constraint is active even though the current solution strictly falls into the feasible domain. In other words, it may happen that a constraint cuts the feasible domain directing the algorithm towards a very conservative solution. This is clearly shown in Figure 2 of Levy and Lavan 2006 \cite{lavan2006optimal}. In the SLP algorithm used in this work, these undesired constraints are disregarded and their effect is nullified in the following optimization iterations.\\

\noindent \textbf{Continuation scheme for parameter control}.
The optimization problem \eqref{eq:optprob2} includes several highly nonlinear components, namely the differential equivalents of the max functions in the aggregated constraint. Therefore, difficulties to converge smoothly towards a good optimized solution are expected. A common approach for promoting a smooth convergence of the optimization process is to gradually increase the parameters that control the degree of nonlinearity. This applies to the parameters $p$ and $q$ in Eq. \eqref{eq:constr2} and Eq. \eqref{eq:constr3}. Furthermore, a conservative move-limit strategy is applied in the solution of the sub-problems, meaning that in each optimization iteration $i$ the updates of $\textbf{x}$ are searched in a close neighborhood of the solution corresponding to the previous iteration $i-1$: $\textbf{x}_{i-1} - ml \leq \textbf{x}_{i} \leq \textbf{x}_{i-1} + ml$. Specific details regarding the values of these parameters are given in the numerical examples of Sec. \ref{sec:numex}. \\

\noindent \textbf{Convergence criteria}.
The methodology is assumed to have reached the final solution in a $k$-th optimization sub-problem \eqref{eq:optprob2} after a minimum of $i_{min}$ iterations, and once we have that: $\Delta \textbf{x} < \delta$, with $\Delta \textbf{x} = \norm{\textbf{x}_{i}-\textbf{x}_{i-1}}$ and $\delta=0.10\,ml\,\sqrt{N_{d}}\,$. The value of $\delta$ and $i_{min}$ considered is given in Sec. \ref{sec:numex}.
The overall optimization approach halts once the final solution $\textbf{x}^{*,k}$ of the current sub-problem considered \eqref{eq:optprob2} satisfies the drift constraints in all the failure scenarios identified for the problem. The solution $\textbf{x}^{*,k}$ is hence considered the final solution of \eqref{eq:optprob}.


\section{Numerical examples}
\label{sec:numex}
In the following section, several numerical results are presented and discussed. They are obtained by optimizing two realistic structures. As already mentioned in Sec. \ref{sec:sensanal}, the optimization problem formulation \eqref{eq:optprob2} is solved with a modified SLP approach inspired by the cutting planes method, which has been implemented in Python 2.7 by the author. 
All the numerical analyses were performed on a Linux machine with 8 Gb of RAM and a dual core Intel i7 CPU at 2.00 GHz.

\begin{figure*}[htbp!]
\fontsize{9}{11}\selectfont 
\centering
\begin{subfigure}[t]{0.49\textwidth}
  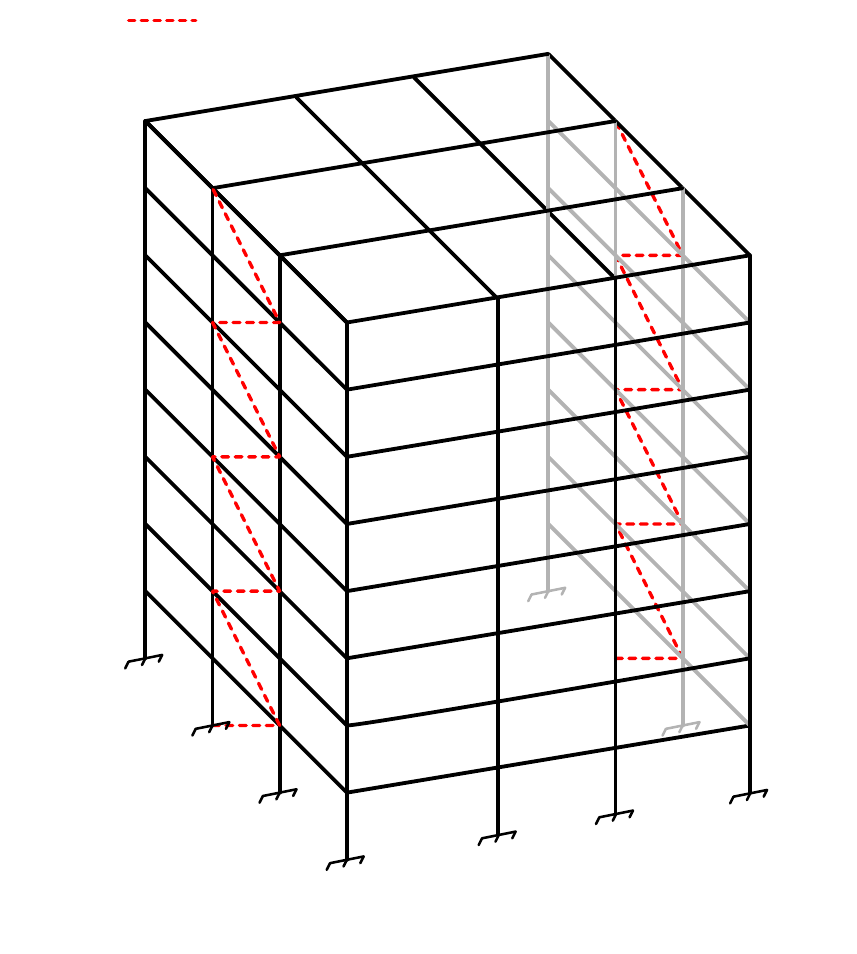
  \caption{Example 1}
  \label{fig:ex1a}
  \end{subfigure}
  ~ 
  \begin{subfigure}[t]{0.49\textwidth}
  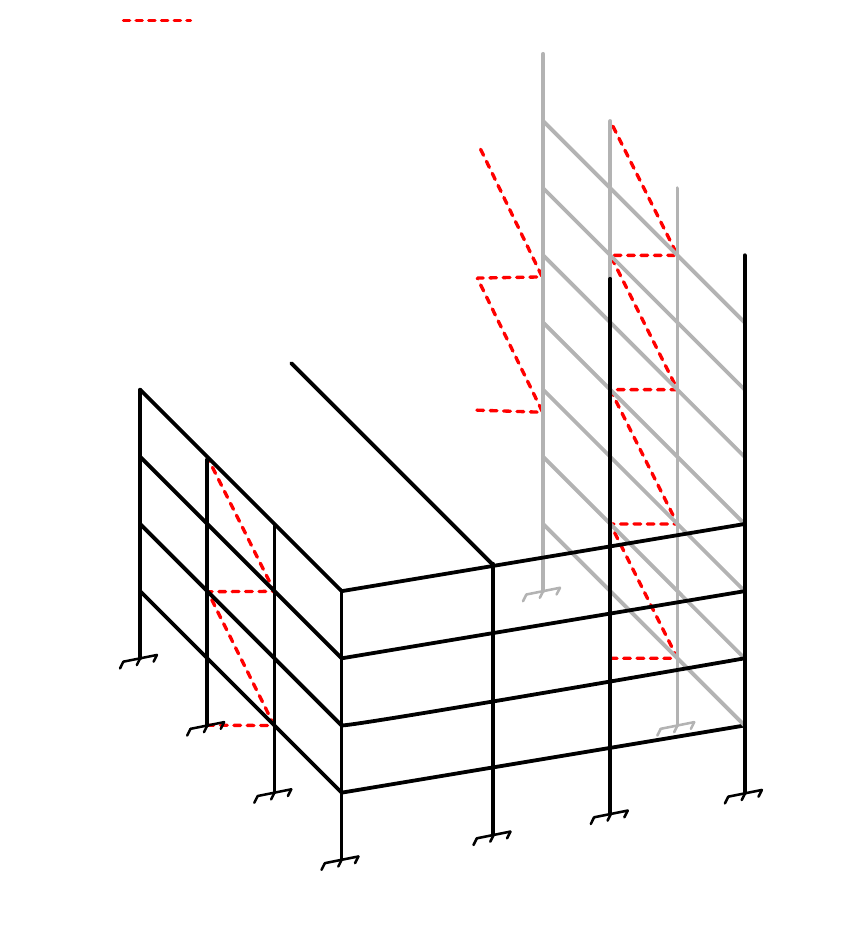
  \caption{Example 2}
  \label{fig:ex1b}
  \end{subfigure}
\caption{Asymmetric structures considered in Sec. \ref{subsec:numex1} (left) and Sec. \ref{subsec:numex2} (right)}     
\label{fig:ex1}  
\end{figure*}

We consider two examples of asymmetric frames made of reinforced concrete, as introduced in Tso and Yao in 1994 \cite{tso1994seismic}. These two test cases were also considered by Lavan and Levy in 2006 \cite{lavan2006optimalper} where an optimal continuous damping was found, and in Lavan and Amir in 2014 \cite{lavan2014simultaneous} but yielding a discrete damping distribution. 
The same examples where also considered by Pollini et al. in 2016 \cite{pollini2016towards}, where a realistic retrofitting cost function was minimized. 
In both examples the column sizes are 0.5m $\times$ 0.5m in frames 1 and 2; 0.7m $\times$ 0.7m in frames 3 and 4 (see Figure \ref{fig:ex1}). 
The beam sizes are 0.4m $\times$ 0.6m and the floor mass is uniformly distributed with a weight of 0.75 ton/m$^{2}$. Regarding the ground motion acceleration, out of the ensemble LA 10\% in 50 years \cite{la1050}, LA16 has the largest maximal displacement for reasonable values of the periods of the structures in both examples. 
Hence LA16 was the ground motion considered first in both examples, acting in the $y$ direction \cite{lavan2006optimalper}. 
In the present work, we consider 5\% of critical damping for the first two modes in order to build the Rayleigh damping matrix of the structures.

In each example, $16$ dampers can potentially be sized and placed in the structures considered.
Three groups of failure scenarios are considered at the same time: \\ 
1) In the first, no failure is considered. 
The dampers are optimized without considering any damage scenario; \\
2) In the second, complete failure of one damper at a time is considered. 
This is equivalent to group $16$ dampers ($n_{c}=16$) into groups of $1$ damper ($k_{c}=1$) per group. 
Thus, the number of distinct complete failure scenarios is:
\begin{equation}
\begin{split}
 &\frac{n_{c}!}{k_{c}!(n_{c}-k_{c})!} \quad  \text{with} \quad n_{c}=16,\; k_{c}=1\\
 & \text{Hence:} \quad  \frac{16!}{1!(15)!}=16
\end{split}
\end{equation}
Numerically, the complete damage of a $i$-th damper in the failure scenario $\alpha$ is enforced by multiplying the corresponding damping coefficient by $0$: $c_{d,i}^{\alpha}= 0 \times c_{d,i}$ with $\alpha=1,2,\dots,16$; \\ 
3) In the third, partial failure of two dampers at a time is considered.
This is equivalent to group $16$ dampers ($n_{p}=16$) into groups of $2$ dampers ($k_{p}=2$) per group. 
Thus, the number of distinct partial failure scenarios is:
\begin{equation}
\begin{split}
 &\frac{n_{p}!}{k_{p}!(n_{p}-k_{p})!} \quad  \text{with} \quad n_{p}=16,\; k_{p}=2\\
 & \text{Hence:} \quad  \frac{16!}{2!(14)!}=120
\end{split}
\end{equation}
Numerically, the partial damage of a $i$-th damper in the failure scenario $\alpha$ is enforced by multiplying the corresponding damping coefficient by $0.5$ ($50\%$ of damage): $c_{d,i}^{\alpha}= 0.5 \times c_{d,i}$ with $\alpha=1,2,\dots,120$.

Therefore, the total number of failure scenarios considered in the following numerical examples is $N_{FS}=1+16+120=137$.
In principle many more failure scenarios can be identified, by varying the number of dampers simultaneously damaged and the level of damage. 
However, as it will be shown in the numerical examples, only few of the scenarios previously identified will be actually governing the design. 

\normalsize

In regards to the parameters that define the approach discussed herein, the following settings were selected after numerical experiments:
the maximum damping coefficient available from Eq. \eqref{eq:cdxi} is $\bar{c}_{d}=150000 \frac{kNs}{m}$;
the maximum allowed value of inter-story drift introduced in Eq. \eqref{eq:constr1} is $d_{allow}=3.5$ cm, i.e. $1\%$ of
story height; the parameters $p$ and $q$ introduced in Eq. \eqref{eq:constr2} and Eq. \eqref{eq:constr3} are set to $100$ and increased by steps of $500$ up to $10^{6}$; the parameter $\epsilon$ used in Eq. \eqref{eq:firstws} is set to $0.05$; the moving limit considered is $ml=0.02$; the value of $\delta$ considered for the convergence criteria is $0.008$, and $i_{min}=50$ iterations.


\subsection{Example 1: Eight-story three bay by three bay asymmetric structure}
\label{subsec:numex1}
A 3-D view of the first frame to be optimized is displayed in Figure \ref{fig:ex1a}. Based on the results of Lavan and Levy \cite{lavan2006optimalper}, 16 potential locations for dampers are assigned at the exterior frames in the $y$ direction.
As a result of the working-set strategy adopted, four optimization analyses were performed, considering sub-problems with $1$, $2$, $3$ and $4$ failure scenarios respectively. The optimization analyses run for $82$, $105$, $91$ and $75$ iterations respectively, for a total computational time of $15$ min and $8$ s over $353$ iterations.
The final optimized solution is shown in Table \ref{tab:ex1}. 
For comparison, the results obtained without considering any failure scenarios are also included in the table.
They are referred to as ``basic design'' in contrast to the ``fail-safe design''.
It can be observed that the two solutions are significantly different in terms of dampers' number and size. 
The basic design has 26\% of the total added damping of the fail-safe design.
Moreover, the fail-safe design relies on more dampers and of larger size. 
In particular, the fail-safe design has $13$ dampers and the final value of the objective function is $J=615\,875$ kNs/m.
The basic design has $9$ dampers and the final value of the objective function is $J=161\,925$ kNs/m

\begin{table}[htbp!]
  \centering
  \caption{Optimized damping values for the asymmetric eight-story frame of Sec. \ref{subsec:numex1}. 
  The results obtained with the working-set strategy and with the full set of failure scenarios are both presented.
  The fail-safe results are shown together with the results obtained without considering any damage scenarios (i.e. the basic design), for comparison. The fail-safe design has $13$ dampers instead of $9$, as in the basic design. The dampers' sizes are in general larger in the fail-safe case. In the last row, the final values of the objective cost functions are shown. All results have been obtained considering the record LA16}
    \begin{tabular}{rrrr}
    \toprule
    Location & Basic design & Fail-safe design & Fail-safe design\\
     & & (Working set) &  (Full set)\\
     & [kNs/m] & [kNs/m] & [kNs/m] \\
    \midrule
    1     & 1\,682  & 126\,469 & 126\,470 \\
    2     & 32\,585 & 61\,554 & 61\,389 \\
    3     & 23\,454 & 96\,247 & 96\,367 \\
    4     & 19\,285 & 51\,637 & 51\,839 \\
    5     & 13\,054 & 28\,122 & 28\,432 \\
    6     & 0     & 24\,871 & 24\,706 \\
    7     & 0     & 20\,444 & 20\,001 \\
    8     & 0     & 0  & 0\\
    9     & 0     & 67\,795 & 68\,101 \\
    10    & 24\,584 & 25\,045 & 24\,908 \\
    11    & 29\,129 & 32\,975 & 32\,828 \\
    12    & 17\,490 & 46\,308 & 46\,509 \\
    13    & 662   & 13\,264 & 13\,335\\
    14    & 0     & 21\,144  & 20\,979\\
    15-16    & 0     & 0 & 0\\
    \midrule
    $J$    & 161\,925     & 615\,875 & 615\,864 \\
    \bottomrule
    \end{tabular}%
  \label{tab:ex1}%
\end{table}%

\begin{figure}[htbp!]
\centering
  \includegraphics[width=.8\textwidth]{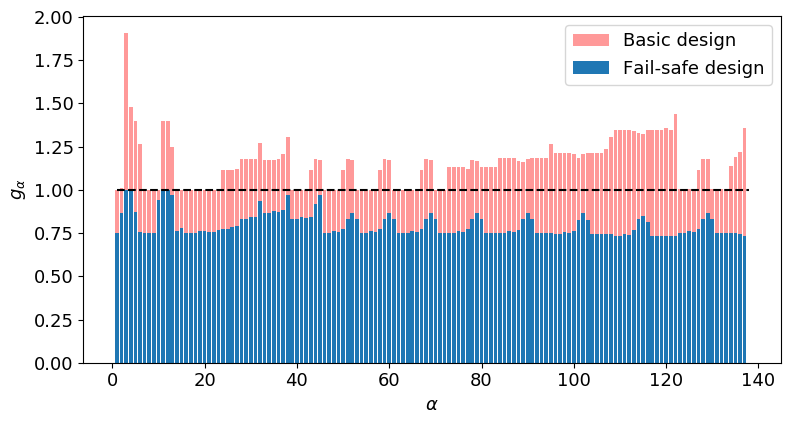}
\caption{Maximum value of the drift constraints $g_{\alpha}$ of the retrofitted structure of Example 1 (Sec. \ref{subsec:numex1}) for all failure scenarios. Results for the fail-safe design (blue) and basic design (red). The record considered for optimization is LA16. The red dashed line marks the maximum value allowed of normalized inter-story drift, i.e. $1.0$. The basic design significantly violates the inter-story drift constraint in several failure scenarios}
\label{fig:dcfsasymm}
\end{figure}
\begin{figure}[htbp!]
\centering
  \includegraphics[width=.7\textwidth]{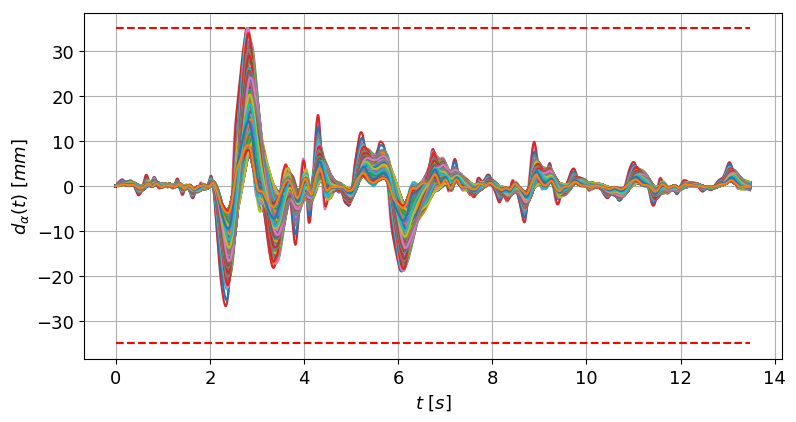}
\caption{Values in time of the inter-story drifts $\text{d}(t) = \text{H}\text{u}(t)$ of the structure equipped with the fail-safe optimized dampers' layout for all fail-safe scenarios, in Example 1 (Sec. \ref{subsec:numex1}). The record considered for optimization is LA16. The red dashed lines mark the maximum allowed inter-story drift $d_{allow}=35$ mm}
\label{fig:dctfsasymm}
\end{figure}

The structure has been tested with both the fail-safe and basic dampers' designs for all $137$ failure scenarios, to compare the performances of the two solutions.
Figure \eqref{fig:dcfsasymm} shows a plot of the maximum value of drift constraint (Eq. \eqref{eq:constr4}) for all failure cases. 
It is possible to observe that with the fail-safe design in correspondence of the the optimized solution few failure scenarios are actually governing the design because their associated normalized peak drift is equal to, or close to, one.
With the basic design, instead, a significant constraint violation is observed for several failure scenarios.
This highlights the superior performance and safety level of the layout of dampers obtained with the fail-safe approach discussed in this paper.
Figure \eqref{fig:dctfsasymm} shows the time-history of all the inter-story drifts of the structure retrofitted with the fail-safe dampers' design for all failure scenarios defined. 
None of the inter-story drifts exceeds the maximum allowed value. 

The optimized fail-safe design was evaluated with the other 19 ground motions in the ensemble, and no other constraint violations were encountered for all the failure scenarios.

To verify the extent of the computational saving achieved using the working-set strategy, we performed an additional optimization analysis considering at once all the failure scenarios.
The purpose of this test was to compare the computational cost required for the solution of the full problem \eqref{eq:optprob} to the one required for the solution of the sub-problems \eqref{eq:optprob2}, as explained in Sec. \ref{sec:workset}. The optimization converged after $71$ iteration, taking $11$ h $36$ min $19$ s.
\begin{table}[htbp!]
  \centering
  \caption{Comparison of the computational cost required by the working-set strategy and by the full problem in Example 1 (Sec. \ref{subsec:numex1}). The number of function evaluations counts the number of time-history and adjoint sensitivity analyses performed. The working-set approach requires a number of function evaluations ten times smaller}
   \begin{tabular}{lllll}
    \toprule
     Approach & \multicolumn{1}{l}{Iterations} & \multicolumn{1}{l}{Number of failure} & \multicolumn{1}{l}{Time} & \multicolumn{1}{l}{Number of function} \\
     &  & \multicolumn{1}{l}{scenarios} &  & \multicolumn{1}{l}{evaluations} \\
     \midrule
    Working set & $\{82,105,91,75 \}$ & $\{1,2,3,4 \}$ & $15.14$ min& $1\,730$\\
    Full set & $\{71\}$ & $\{137\}$& $696.31$ min & $19\,454$\\
        \bottomrule
    \end{tabular}%
  \label{tab:wsfull}%
\end{table}%
The final results are presented in Table \ref{tab:ex1}.
In Table \ref{tab:wsfull} we compare the computational time and effort of the working-set approach with the one required for the full problem. The working-set approach requires the solution of multiple sub-problems, which are essentially a relaxation of the original full problem. However, every sub-problem considers a significantly smaller number of failure scenarios. This results is a smaller number of constraints to be considered in every sub-problem, and hence less time-history and adjoint sensitivity analyses.
As a result, the computational time is reduced of the $97.8\%$, and only approximately a tenth of the function evaluations (i.e time-history and sensitivity analyses) is required.


\subsection{Example 2: Eight-story three bay by three bay setback frame structure}
\label{subsec:numex2}
A 3-D view of the second frame considered is shown in Figure \ref{fig:ex1b}. 
Also in this case, based on the results of Lavan and Levy \cite{lavan2006optimalper}, 16 potential locations for dampers are assigned at the exterior frames in the $y$ direction.
As a result of the working-set strategy adopted, five optimization analyses were performed. 
Each consisted of a sub-problem with $1$, $2$, $3$, $4$ and $9$ failure scenarios respectively. 
The optimization analyses run for $68$, $60$, $72$, $74$ and $67$ iterations respectively, for a total computational time of $21$ min and $35$ s over $341$ iterations.
The final optimized solution is shown in Table \ref{tab:ex2}. 
Also in this case, the dampers optimized without considering any failure scenarios are included in the table for comparison (i.e. basic design).
They are referred to as ``basic design'' in contrast to the ``fail-safe design''.
It can be observed that also in this case the two solutions are significantly different in terms of dampers' number and size. 
The basic design has 43\% of the total added damping of the fail-safe design.
In fact, the fail-safe design relies on more dampers and of larger size. 
In particular, the fail-safe design has $11$ dampers and the final value of the objective function is $J=153\,138$ kNs/m.
The basic design has $4$ dampers and the final value of the objective function is $J=66\,252$ kNs/m.

\begin{table}[htbp!]
  \centering
  \caption{Optimization results of the asymmetric eight-story setback frame of Sec. \ref{subsec:numex2}. 
  The results obtained with the working-set strategy considering only the record LA16, and the records LA14, LA16, and LA18 at the same time are both listed.
  The fail-safe results are shown together with the results obtained without considering any damage scenarios (i.e. the basic design), for comparison. It can be observed that the fail-safe design involves $11$ dampers instead of $4$, as in the basic design. The dampers' sizes are in general larger in the fail-safe case. In the last row, the final values of the objective functions are listed}
   \begin{tabular}{rrrr}
    \toprule
    Location & Basic design & Fail-safe design & Fail-safe design \\
    & (LA16) & (LA16) & (LA14, LA16, LA18)\\
     & [kNs/m] & [kNs/m] & [kNs/m] \\
    \midrule
    1 & 0     & 18\,304 & 19\,890  \\
    2 & 14\,996 & 8\,963 & 9\,370 \\
    3 & 0     & 10\,965 & 13\,726 \\
    4 & 0     & 0 & 126\\
    5 & 0     & 7\,613 & 5\,237 \\
    6 & 0     & 5\,661 & 6\,236 \\
    7-8 & 0     & 0 & 0 \\
    9 & 0     & 15\,508 & 13\,572 \\
    10 & 22\,198 & 20\,947 & 21\,726 \\
    11 & 26\,552 & 34\,011 & 36\,235 \\
    12 & 2\,506  & 25\,812 & 26\,569 \\
    13 & 0     & 4\,129 & 4\,741 \\
    14 & 0     & 1\,225 & 0 \\
    15-16 & 0     & 0 & 0 \\
    \midrule
    $J$ & 66\,252 & 153\,138 & 156\,428 \\
    \bottomrule
    \end{tabular}%
  \label{tab:ex2}%
\end{table}%

\begin{figure}[htbp!]
\centering
  \includegraphics[width=.8\textwidth]{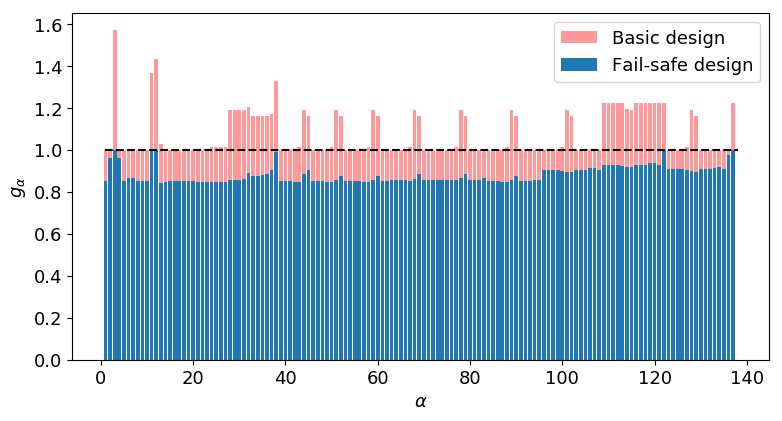}
\caption{Maximum value of the drift constraints $g_{\alpha}$ of the retrofitted structure of Example 2 (Sec. \ref{subsec:numex2}) for all failure scenarios. Results for the fail-safe design (blue) and basic design (red). The record considered for optimization is LA16. The red dashed line marks the maximum value allowed of normalized inter-story drift, i.e. $1.0$. The basic design significantly violates the inter-story drift constraint in several failure scenarios}
\label{fig:dcfsasymmset}
\end{figure}
\begin{figure}[htbp!]
\centering
  \includegraphics[width=.7\textwidth]{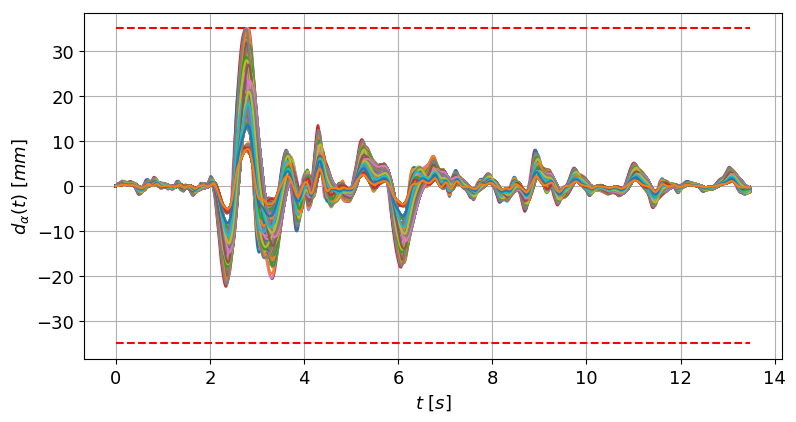}
\caption{Values in time of the inter-story drifts $\text{d}(t) = \text{H}\text{u}(t)$ of the structure with optimized dampers' layout for all fail-safe scenarios, in Example 2 (Sec. \ref{subsec:numex2}). The record considered for optimization is LA16. The red dashed lines mark the maximum allowed inter-story drift $d_{allow}=0.035m$}
\label{fig:dctfsasymmset}
\end{figure}

Also in this case, the structure has been tested with both the fail-safe and basic dampers' designs for all $137$ failure scenarios, to compare the performances of the two solutions.
Figure \eqref{fig:dcfsasymmset} shows a plot of the maximum value of drift constraint (Eq. \eqref{eq:constr4}) for all failure cases. 
It is possible to observe that with the fail-safe design in correspondence of the the optimized solution few failure scenarios are actually governing the design because their associated normalized peak drift is equal to, or close to, one.
With the basic design, instead, a significant constraint violation is observed for several failure scenarios.
This highlights the superior performance and safety level of the layout of dampers obtained with the fail-safe approach discussed in this paper.
Figure \eqref{fig:dctfsasymmset} shows the values in time of all the inter-story drifts of the structure retrofitted with the fail-safe layout of dampers for all the failure scenarios defined. 
None of the inter-story drifts exceeds the maximum allowed value.

\begin{figure}[h]
\centering
  \includegraphics[width=.8\textwidth]{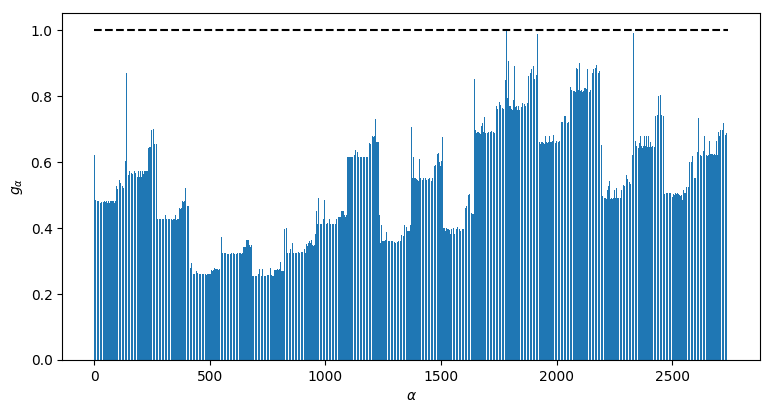}
\caption{Maximum value of the drift constraints $g_{\alpha}$ of the structure in Example 2 (Sec. \ref{subsec:numex2}), with the fail-safe layout of dampers optimized considering the records LA14, LA16, and LA18. The plotted results consider all fail-safe scenarios and all the records from the LA 10\% in 50 years ensemble. The red dashed line marks the maximum value allowed of normalized inter-story drift, i.e. $1.0$}
\label{fig:dcfsasymmsetens}
\end{figure}

The optimized fail-safe design was tested with the other 19 ground motions in the ensemble.
In two cases a significant constraint violation was observed: $7\%$ with LA14  and $4\%$ with LA18.
Thus another optimization analysis was performed considering three records at the same time, namely LA14, LA16, and LA18.
The results are shown in Table \ref{tab:ex2}. 
The optimization analysis run through three sub-problems for $60$, $69$, and $74$ iterations, for a total of $203$ iterations and a  computational time  of $1$ h $47$ min $51$ s.
Each sub-problem consisted of $3$, $4$, and $5$ failure scenarios.
The obtained fail-safe design was then tested with all the records from the ensemble considered, and no constraint violation was encountered for all failure scenarios. This is shown in Figure \ref{fig:dcfsasymmsetens}.


The numerical results from Sec. \ref{subsec:numex1} and Sec. \ref{subsec:numex2} show that significantly different optimized solutions are obtained if failure scenarios are considered in the design phase.
This is clearly shown in Tables \ref{tab:ex1} and \ref{tab:ex2}.
The solutions significantly differ from the solutions obtained without considering any damage in the dampers in terms of number of dampers allocated in the structure and dampers' size.
Moreover, the designs obtained with the fail-safe approach show superior performance and safety levels compared to the designs optimized without considering any damage in the dampers. This can be observed in Figure \ref{fig:dcfsasymm} and Figure \ref{fig:dcfsasymmset}. 
In the first example, we also showed that by using a working-set strategy it is possible to significantly reduce of ten times the computational cost required for the optimization process.
This is a crucial aspect, because it allows the approach discussed herein to be implemented on standard desktop computers, and hence to assist in their activity engineers and practitioners that have access only to modest computational resources.
In the second example, the fail-safe design obtained considering a single dominant acceleration record (i.e. LA16) did not fulfil the inter-story drift requirements with two of the records from the LA 10\% in 50 years ensemble.
Thus an additional optimization analysis was performed considering simultaneously three records: LA14, LA16, and LA18.
In this way it was shown that the methodology handles realistic ensembles of ground motions, and it can thus be used in practical performance-based design applications of 3-D irregular structures.
Moreover, thanks to the working-set strategy adopted, the computational effort proved reasonable also when three acceleration records were considered at the same time. Computational cost that would have been most likely prohibitive otherwise.

\section{Final considerations}
\label{sec:final}
This paper presents a new optimization approach for designing minimum-cost fail-safe distributions of fluid viscous dampers for seismic retrofitting. 
Failure is modeled as either complete damage of the dampers or degradation of the dampers' properties. 
The dampers' cost function is minimized, with constrains on the inter-story drifts at the peripheries of irregular 3-D structures. 
These are computed with time-history analyses considering an ensemble of realistic ground motions. 
Therefore, the proposed methodology can be used for the fail-safe performance-based seismic retrofitting of 3-D irregular structures.

The novelty of the proposed approach lies in its formulation, which allows to optimize an added damping system with superior performance and safety levels compared to other similar approaches available in the literature. The computational cost is significantly reduced by means of a working-set strategy: only few dominant failure scenarios are actually considered during the optimization, and the final designs fulfill all the performance constraints associated to all the failure scenarios initially identified.

The numerical results show how the proposed methodology successfully handled realistic design cases. A large number of failure scenarios was considered successfully, solving a sequence of relaxed sub-problems instead of the original full problem.
Each sub-problem considered only a working-set of constraints associated to the most critical failure scenarios.
The methodology discussed herein ensures that the working-set is updated after every sub-problem is solved, and that the failure scenarios previously considered are contained in the new working-set. In other words, the working-sets expand until the overall optimization process terminates.
The numerical results also show that the optimized dampers' layout and size can be significantly different when failure scenarios are considered in the design phase: the algorithm tends to place more dampers and of bigger size. 
This is in good agreement with the engineering intuition according to which if failure scenarios are considered in the design, the level of redundancy of the designed system increases.
Moreover, the working-set strategy significantly reduces the computational effort required for optimization. 
This means that the methodology can be used by engineers in practice relying on standard computational resources.
The methodology is expected to promote the fail-safe optimization-based design of dampers even for large scale structures, where the number of design variables may become very large and other optimization approaches (e.g. genetic algorithms) would require prohibitive computational efforts and resources.

\section*{Acknowledgments}
The author wishes to thank Assoc. Prof. Oren Lavan from the Technion - Israel Institute of Technology for his valuable feedback on the final version of the paper.

\bibliographystyle{unsrt}


\end{document}